\documentclass[apj]{emulateapj}
\usepackage{epsfig}
\usepackage{wrapfig,setspace}
\usepackage{xcolor}

\newcommand{\etal}{et al.}

\def\simlt{\mathrel{\hbox{\rlap{\hbox{\lower4pt\hbox{$\sim$}}}\hbox{$<$}}}}
\def\simgt{\mathrel{\hbox{\rlap{\hbox{\lower4pt\hbox{$\sim$}}}\hbox{$>$}}}}

\def\chandra{{\it Chandra}}
\def\xmm{{\it XMM-Newton}}

\def\nicer{{\it NICER\/}}

\def\psr{PSR~J1210$-$5226}
\def\pks{PKS~1209$-$51/52}

\def\simlt{\mathrel{\hbox{\rlap{\hbox{\lower4pt\hbox{$\sim$}}}\hbox{$<$}}}}
\def\simgt{\mathrel{\hbox{\rlap{\hbox{\lower4pt\hbox{$\sim$}}}\hbox{$>$}}}}
\def\cco{1E~1207.4$-$5209}

\slugcomment{}
 
\shorttitle{Timing Behavior of \cco}
\shortauthors{Gotthelf \& Halpern}

\begin{document}

\title{
The Timing Behavior of the Central Compact Object Pulsar \cco\
}

\author{E. V. Gotthelf and J. P. Halpern}

\affil{Columbia Astrophysics Laboratory, Columbia University,
New York, NY 10027-6601, USA; eric@astro.columbia.edu}

\begin{abstract}
  We present 20 years of timing observations for \cco, the central
  compact object in supernova remnant \pks, to follow up on our
  detection of an unexpected timing glitch in its spin-down.  Using
  new \xmm\ and \nicer\ observations of \cco, we now find that the
  phase ephemeris can be well-modelled by either two small glitches,
  or extreme timing noise. The implied magnitudes of the frequency
  glitches are $\Delta f/f = (9\pm2)\times10^{-10}$ and $\Delta f/f =
  (3.7\pm0.7)\times10^{-10}$, at epochs 2010.9 and 2014.4,
  respectively. The updated timing solutions also rule out our
  previous suggestion of a large glitch in the frequency derivative
  $\dot f$.  No other canonical pulsar with such a small spin-down
  rate ($\dot f = -1.2 \times 10^{-16}$~Hz~s$^{-1}$) or surface dipole
  magnetic field strength ($B_s = 9.8\times 10^{10}$~G) has been
  observed to glitch; the glitch activity parameter of \cco\ is larger
  than that of more energetic pulsars.  Alternative parameterizations
  that do not involve glitches can fit the data, but they have timing
  residuals or a second frequency derivative $\ddot f$ that are orders
  of magnitude larger than in pulsars with similar spin-down
  parameters.  These timing properties of \cco\ further motivate the
  leading theory of central compact objects, that an initial $B$-field
  of normal strength was buried in the neutron star crust by fallback
  of supernova ejecta, suppressing the surface dipole field.  The slow
  reemergence of the buried field may be involved in triggering
  glitches or excess timing noise.
  
\end{abstract}

\keywords{pulsars: individual (\cco, \psr) --- stars: neutron}

\section {Introduction}

The central compact object (CCO) \cco\ in the supernova remnant (SNR)
\pks\ has been studied intensively because of its unusual timing and
spectral properties.  It was the first CCO pulsar discovered
\citep{zav00}, the first isolated neutron star (NS) to display strong
absorption lines in its X-ray spectrum
\citep{san02,mer02,big03,del04}, and most recently, the first CCO to
show glitch activity \citep{got18}.  \cco\ is one of the three known
CCO pulsars, all with characteristic weak surface dipole magnetic
field strength, ($2.9,3.1,$ and $9.8$)$\times10^{10}$~G, the smallest
known among young pulsars \citep{got07,got13,hal10,hal11,hal15}.

CCOs are young NSs associated with SNRs defined by their steady
surface thermal X-ray emission, lack of surrounding pulsar wind
nebula, and nondetection at any other wavelength (\citealt{pav02}; see
\citealt{del17} for a recent review). CCOs are as numerous as other
classes of NS in SNRs, implying that they represent a significant
fraction of NS births.  In addition to the three CCO pulsars, $\sim7$
NSs with similar properties have eluded searches for pulsations.  They
may have even weaker magnetic fields, more uniform surface temperature
distribution, or an unfavorable viewing geometry.

The spin-down magnetic field inferred for \cco, $B_s =
9.8\times10^{10}$~G, is remarkably close to
$B\approx8\times10^{10}$~G, the value measured from its spectroscopic
absorption features, interpreted as the electron cyclotron fundamental
at 0.7~keV and its harmonics.  This agreement has all but eliminated
competing ideas for the origin of the absorption lines, and provides a
convincing confirmation of the surface $B$-field strength.

The recent discovery of a glitch from \cco\ \citep{got18}
is most unexpected given the absence of glitches in pulsars with
such small $\dot f$ and $B_s$.  Possibly related is the problem of how
hot spots are created on the NS surface, as evidenced by the X-ray
pulse modulation, in the absence of a strong magnetic field.
\citet{hal10} reviewed theoretical arguments for CCOs having magnetar
strength internal toroidal fields $\sim10^{14}$~G, possibly buried during
the formation of the NS, that could account for their hot spots and
high X-ray luminosity without contributing to their weak external dipole
fields.  \cite{ho15} hypothesized that glitch activity in CCOs may be
triggered by such strong magnetic fields diffusing through the NS crust
and interacting with the neutron superfluid there.

We present \chandra, \xmm\, and \nicer\ observations of \cco\ 
that confirm the original detection of a glitch and reveal a
second glitch, and a possibly third, that suggest a recurrence time
of 4--10 years.
In Section~\ref{sec:data}, we describe the new X-ray timing observations.
In Section~\ref{sec:timing}, we present the updated 
timing solutions that rule out a large glitch in the
frequency derivative as previously reported. We also explore
alternative timing models for the pulsar rotation that can be
interpreted as timing noise. Section~\ref{sec:discussion} compares
the results with the general pulsar population,
and Section~\ref{sec:conclusions} concludes with implications for
the origin of glitches, timing noise, and CCOs themselves.

\section {Data Analysis}
\label{sec:data}

We have obtained new timing observations of \cco\ using the \nicer\
and \xmm\ observatories that allow us to resolve ambiguities in the
glitch analysis reported in \cite{got18} and to consider alternative
interpretations.  We supplement archival \nicer\ data sets starting
from 2017 July 24 with our subsequent AO1 guest
observer data. We include in this work two new \xmm\ AO18 observations
obtained as part of our semi-annual monitoring program.  Because of the
increase in the low-energy X-ray opacity of the \chandra\ ACIS window,
it is no longer practical to use that instrument given the soft
X-ray spectrum of \cco.

Table~\ref{obslog} presents a  complete log of timing observations for \cco.  
Previously published \chandra\ and \xmm\ data
sets used herein are fully described in our earlier work
\citep{got07,hal11,got13,hal15}. Below we detail the preparation of
the \nicer\ data sets, included for the first time in our analysis of
the pulsar. All data sets were reprocessed and reduced using the
latest software for each mission. Photon arrival times were converted
to barycentric dynamical time (TDB) using the DE405 solar system
ephemeris and the \chandra\ coordinates given in \citet{got13}, shown in
Table~\ref{tab:ephemone}.  Significant proper motion has not been
detected \citep{hal15}.  In this analysis we include only photons
that fall in the 0.5$-$1.6~keV energy range, optimal for the pulsar's
observed spectrum. For the \xmm\ and the \chandra\ data sets, we
extracted photons using $30^{\prime\prime}$ and $1\farcs8$ radius
circular apertures, respectively.

The Neutron Star Interior Composition Explorer (\nicer;
\citealt{gen17}) is an X-ray telescope attached to the International
Space Station that provides sub microsecond time resolution in the
0.2$-$12 ~keV band.  The \nicer\ telescope consists of a set of 52
operational non-imaging silicon drift
detectors \citep{pri16}, each at the focus of an X-ray concentrator
\citep{oka16} that subtends a $4^{\prime}$ radius field-of-view. The
nominal effective area of the telescope is 1900~cm$^{2}$ at 1.5~keV.

The \nicer\ data sets were reduced and analyzed using the NICERDAD
software suite distributed in the FTOOLS package, version
24Jun2019\_V6.26.1, and the most up-to-date calibration files. We
generated cleaned event files using the {\tt nicerl2} script that
applied the standard filtering criteria. The data were further reduced
by excluding detectors with anomalous count rates $>5$ sigma above the
mean rate, computed using all available detectors, in the energy range
of interest. Similarly, we iteratively excluded time intervals with
high background rates by comparing the event rate in 10 s steps to the
mean rate.
  
Since its launch in 2017, \nicer\ has observed \cco\ a total of 148
times to-date. Each observation is defined by a unique ObsID number
and typically comprises short exposures (50\% are less than $1.2$~ks)
spread over multiple satellite orbits, and often containing multi-day
gaps within, and between. These exposures are generally too short to
generate a precise pulse phase measurement needed for our timing
analysis. However, by concatenating adjacent observations we obtained eight
sufficiently compact \nicer\ data sets that contained the minimum
number of events required to measure an independent pulse phase, as
described below.

\begin{deluxetable}{llclc}
\tablewidth{0pt}
\tablecolumns{5}
\tablecaption{Log of X-ray Timing Observations of \cco}
\tablehead{
\colhead{Mission}  & \colhead{Instrument}   & \colhead{ObsID}   & \colhead{Date} & \colhead{Expo$^a$} \\
\colhead{} & \colhead{/Mode} & \colhead{} & \colhead{(UT)} & \colhead{(ks)} 
}
\startdata
{\it Chandra\/} & ACIS-S3/CC        & 751            & 2000 Jan\phantom{l} 06   & 32.4  \\
{\it XMM\/}     & EPIC-pn/sw        & 0113050501     & 2001 Dec\phantom{i} 23   & 27.0  \\  
{\it Chandra\/} & ACIS-S/CC         & 2799           & 2002 Jan\phantom{l} 05   & 30.3  \\
{\it XMM\/}     & EPIC-pn/sw        & 0155960301     & 2002 Aug\phantom{} 04    & 128.0 \\
{\it XMM\/}     & EPIC-pn/sw        & 0155960501     & 2002 Aug\phantom{} 06    & 129.0 \\
{\it Chandra\/} & ACIS-S/CC         & 3915           & 2003 Jun\phantom{l} 10   & 155.1 \\
{\it Chandra\/} & ACIS-S/CC         & 4398           & 2003 Jun\phantom{.} 18   & 114.7 \\
{\it XMM\/}     & EPIC-pn/sw        & 0304531501     & 2005 Jun\phantom{.} 22   & 15.1  \\
{\it XMM\/}     & EPIC-pn/sw        & 0304531601     & 2005 Jul\phantom{a} 05   & 18.2  \\
{\it XMM\/}     & EPIC-pn/sw        & 0304531701     & 2005 Jul\phantom{a} 10   & 20.5  \\
{\it XMM\/}     & EPIC-pn/sw        & 0304531801     & 2005 Jul\phantom{a} 11   & 63.4  \\
{\it XMM\/}     & EPIC-pn/sw        & 0304531901     & 2005 Jul\phantom{a} 12   & 14.5  \\
{\it XMM\/}     & EPIC-pn/sw        & 0304532001     & 2005 Jul\phantom{a} 17   & 16.5  \\ 
{\it XMM\/}     & EPIC-pn/sw        & 0304532101     & 2005 Jul\phantom{a} 31   & 17.7  \\
{\it XMM\/}     & EPIC-pn/sw        & 0552810301     & 2008 Jul\phantom{a} 02   & 31.4  \\
{\it XMM\/}     & EPIC-pn/sw        & 0552810401     & 2008 Dec\phantom{i} 22   & 30.4  \\
{\it Chandra\/} & ACIS-S3/CC        &      14199     & 2011 Nov\phantom{} 25    & 31.0  \\
{\it Chandra\/} & ACIS-S3/CC        &      14202     & 2012 Apr\phantom{} 10    & 33.0  \\
{\it XMM\/}     & EPIC-pn/sw        & 0679590101     & 2012 Jun\phantom{.} 22   & 26.5  \\
{\it XMM\/}     & EPIC-pn/sw        & 0679590201     & 2012 Jun\phantom{.} 24   & 22.3  \\
{\it XMM\/}     & EPIC-pn/sw        & 0679590301     & 2012 Jun\phantom{.} 28   & 24.9  \\
{\it XMM\/}     & EPIC-pn/sw        & 0679590401     & 2012 Jul\phantom{a} 02   & 24.5  \\
{\it XMM\/}     & EPIC-pn/sw        & 0679590501     & 2012 Jul\phantom{a} 18   & 27.3  \\
{\it XMM\/}     & EPIC-pn/sw        & 0679590601     & 2012 Aug\phantom{} 11    & 27.3  \\
{\it Chandra\/} & ACIS-S3/CC        &      14200     & 2012 Dec\phantom{i} 01   & 31.1  \\
{\it Chandra\/} & ACIS-S3/CC        &      14203     & 2013 May\phantom{} 19    & 33.0  \\
{\it Chandra\/} & ACIS-S3/CC        &      14201     & 2013 Dec\phantom{i} 04   & 33.0  \\
{\it Chandra\/} & ACIS-S3/CC        &      14204     & 2014 Jun\phantom{.} 20   & 33.0  \\
{\it XMM\/}     & EPIC-pn/sw        &0780000201      & 2016 Jul\phantom{a} 28   & 32.5  \\
{\it XMM\/}     & EPIC-pn/sw        &0800960201      & 2017 Jun\phantom{.} 22   & 33.3  \\
{\it XMM\/}     & EPIC-pn/sw        &0800960301      & 2017 Jun\phantom{.} 23   & 20.7  \\
{\it XMM\/}     & EPIC-pn/sw        &0800960401      & 2017 Jun\phantom{.} 24   & 22.6  \\
{\it XMM\/}     & EPIC-pn/sw        &0800960501      & 2017 Jul\phantom{l} 03   & 23.5  \\
\color{red} {\it NICER\/}$^b$& XTI  &1020270102      & 2017 Jul\phantom{a} 24   & 6.1   \\  
\color{red} {\it NICER\/}$^b$& XTI  &1020270106      & 2017 Jul\phantom{a} 28   & 14.3  \\  
\color{red} {\it NICER\/}$^b$& XTI  &1020270110      & 2017 Aug\phantom{} 01    & 12.1  \\  
{\it XMM\/}     & EPIC-pn/sw        &0800960601      & 2017 Aug\phantom{} 10    & 19.8  \\
{\it Chandra\/} & ACIS-S3/CC        &   19612        & 2017 Oct\phantom{.} 10   & 32.9  \\
\color{red} {\it NICER\/}$^b$& XTI  &1020270130      & 2017 Nov\phantom{} 15    & 20.6  \\  
{\it XMM\/}     & EPIC-pn/sw        &0800960701      & 2017 Dec\phantom{.} 24   & 19.8  \\
{\it XMM\/}     & EPIC-pn/sw        &0821940201      & 2018 Jun\phantom{.} 22   & 33.2  \\ 
{\it Chandra\/} & ACIS-S3/CC        & 19613          & 2018 Aug\phantom{} 27    & 66.3  \\  
\color{red} {\it NICER\/}$^b$ & XTI &1020270153-58   & 2018 Nov\phantom{} 30    & 7.6   \\  
\color{red} {\it XMM\/}$^b$ & EPIC-pn/sw &0821940301 & 2018 Dec\phantom{i} 28   & 26.6  \\
\color{red} {\it NICER\/}$^b$ & XTI &2506010101-02   & 2019 Apr\phantom{l} 04   & 22.3  \\  
\color{red} {\it XMM\/}$^b$ & EPIC-pn/sw &0842280301 & 2019 Jul\phantom{a} 09   & 30.8  \\ 
\color{red} {\it NICER\/}$^b$ & XTI &2506010201-02   & 2019 Jul\phantom{a} 19   & 21.4  \\  
\color{red} {\it NICER\/}$^b$ & XTI &2506010205-13   & 2019 Jul\phantom{a} 26   & 6.7   \\  
\enddata
\tablenotetext{a}{Exposure times for \xmm\ EPIC-pn does not reflect the 29\%  deadtime in the SmallWindow (sw) mode.}
\tablenotetext{b}{Newly reported observations. The \nicer\ data set are denoted by the ObsID and date of the first of the concatenated set of observations (see Section 2 for details).}
\label{obslog}
\end{deluxetable}

\begin{figure}
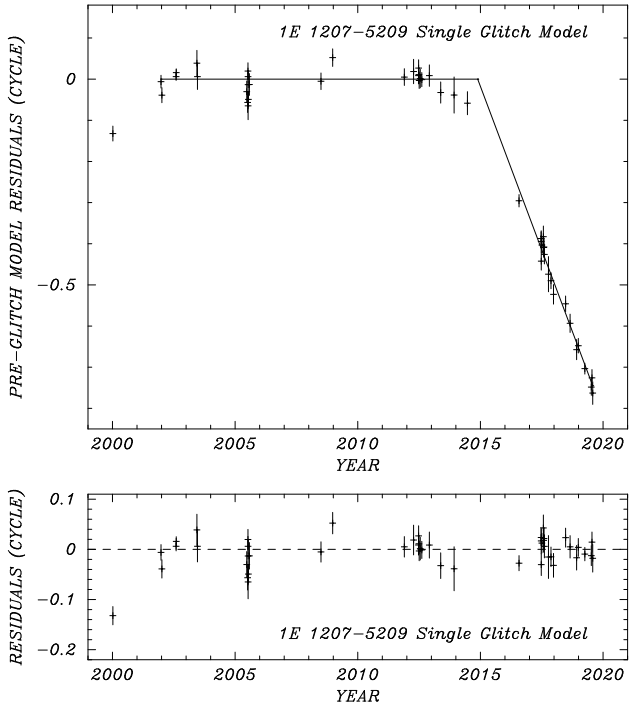

\centerline{\psfig{figure=1e1207_2002_2019_one_glitch.ps,height=0.96\linewidth,angle=270}}
\vspace{2mm}
\centerline{\psfig{figure=1e1207_2002_2019_one_glitch_res.ps,height=0.96\linewidth,angle=270}}
\caption{ 
\footnotesize Top: Pulse-phase residuals from the
  revised pre-glitch timing solution (Table~\ref{tab:ephemone})
  for the single glitch model of \citet{got18}.  A glitch
  epoch of 2014 November~21 is estimated from the intersection of the pre-
  and post-glitch fits (solid lines). The year 2000 \chandra\ data
  point is not included in the fit (see Section~\ref{sec:one} for details).
  Bottom: Combined residuals from the pre- and post-glitch
  timing models. The overall $\chi^2_{\nu} = 1.44$ for 42 degrees of freedom
  (DoF), taking into account the fit parameters for each interval.  
}
\label{fig:oneres}
\end{figure}

\begin{deluxetable}{lc}
\tighten
\tablewidth{0pt}
\tablecolumns{2}
\tablecaption{Single Glitch Ephemerides for \cco}
\tablehead{
\colhead{Parameter} & \colhead{Value$^{\rm a}$ \hfill} 
}
\startdata
R.A. (J2000)                                  & $12^{\rm h}10^{\rm m}00^{\rm s}\!.91$ \\
Decl. (J2000)                                 & $-52^{\circ}26^{\prime}28^{\prime\prime}\!.4$ \\
Surface dipole dipole field, $B_s$            & $9.8 \times 10^{10}$ G\\
Spin-down luminosity, $\dot E$                & $1.1 \times 10^{31}$ erg s$^{-1}$ \\
Characteristic age, $\tau_c\equiv P/2\dot P$  & 303~Myr \\
\cutinhead{Pre-glitch Timing Solution (2002-2014) \hfill}
 Epoch of ephemeris (MJD TDB)              &          54547.00000198           \\                                                                                                                                                                 
 Span of ephemeris (MJD)                   &      52266--56829                   \\
 Frequency, $f$                      &      2.357763492491(28) s$^{-1}$             \\
 Frequency derivative, $\dot f$  &     $-1.2317(66)\times 10^{-16}$ s$^{-2}$  \\
 Period, $P$                            &      0.4241307506816(50) s          \\
 Period derivative, $\dot P$               &      $2.216(12)\times 10^{-17}$     \\
 $\chi^2_{\nu}[{\rm DoF}]$                 &       1.80[25]                     \\
\cutinhead{Post-Glitch Timing Solution (2016-2019) \hfill}  
 Epoch of ephemeris (MJD TDB)              &          58144.00000220           \\
 Span of ephemeris (MJD)                   &      57597--58695                   \\
 Frequency, $f$                      &      2.35776345915(16) s$^{-1}$           \\
 Frequency derivative, $\dot f$  &      $-1.01(12)\times 10^{-16}$ s$^{-2}$    \\
 Period, $P$                           &      0.424130756679(30) s           \\
 Period derivative, $\dot P$               &      $1.81(21)\times 10^{-17}$      \\
 $\chi^2_{\nu}[{\rm DoF}]$                 &      0.91[17]                     \\
\noalign{\vskip 0.5em}\hline\noalign{\vskip 0.5em}
 Glitch epoch (MJD)$^{\rm b}$                &      56982(6)                       \\
 $\Delta f$\                     &      $5.03(16)\times 10^{-9}$  s$^{-1}$    \\
 $\Delta f/f_{\rm pred}$                 &     $2.134(66)\times 10^{ -9}$      \\[-0.5em]
\enddata
\tablenotetext{}{Note -- Derived parameters ($B_s$, $\dot E$,
  $\tau_c$) are based on the pre-glitch timing solution.}
\tablenotetext{a}{Uncertainties in the last digits are given in
  parentheses.}  \tablenotetext{b}{Epoch of the glitch estimated by
  matching the zero phase of the two timing solutions; this assumes a
  constant post-glitch $\dot f$.}
\label{tab:ephemone}
\end{deluxetable}

\begin{figure}
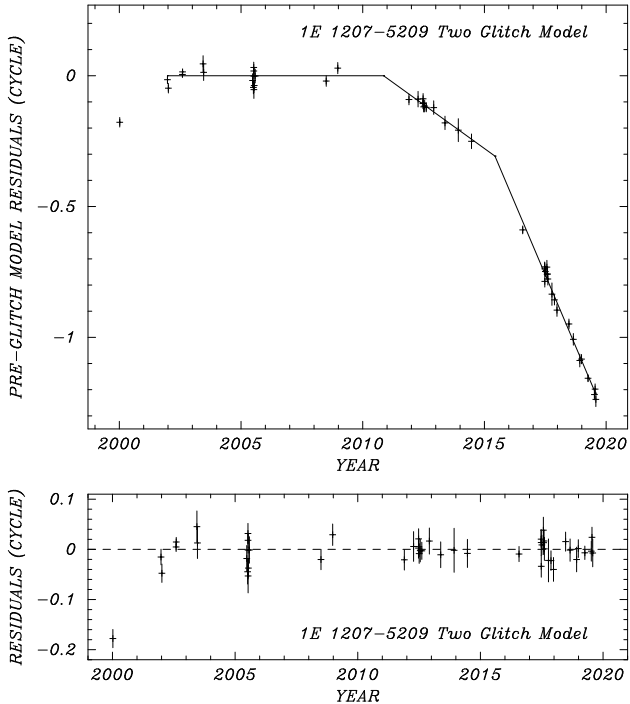

\centerline{\psfig{figure=1e1207_2002_2019_two_glitch.ps,height=0.96\linewidth,angle=270}}
\vspace{2mm}
\centerline{\psfig{figure=1e1207_2002_2019_two_glitch_res.ps,height=0.96\linewidth,angle=270}}
\caption{ \footnotesize Top: Pulse-phase residuals from the
  pre-glitch timing solution presented in Table~\ref{tab:ephem},
  modelled as two successive glitch-like changes in
  frequency.  Glitch epochs of 2010 November~09 and 2014
  May~23 are estimated from the intersection of the respective pre- and
  post-glitch fits (solid lines).  The year 2000 \chandra\ data point
  is not included in the fit (see Section~\ref{sec:two} for details).
  Bottom:
  Combined residuals from timing model fits to the three inter-glitch
  intervals. The overall $\chi^2_{\nu} = 1.23$ for 39 DoF, taking into
  account the fit parameters for each interval. }
\label{fig:postres}
\end{figure}

\begin{deluxetable}{lc}
\tighten
\tablewidth{0pt}
\tablecolumns{2}
\tablecaption{ Two Glitch Ephemerides for \cco}
\tablehead{
\colhead{Parameter} & \colhead{Value$^{\rm a}$ \hfill} 
}
\startdata
\cutinhead{Pre-glitch Timing Solution (2002-2008) \hfill}
 Epoch of ephemeris (MJD TDB)              &          53544.00000442          \\
 Span of ephemeris (MJD)                   &      52266--54822                   \\
 Frequency, $f$                      &      2.357763503102(75) s$^{-1}$             \\
 Frequency derivative, $\dot f$  &   $-1.278(21)\times 10^{-16}$ s$^{-2}$   \\
 Period, $P$                           &      0.424130748773(14) s         \\
 Period derivative, $\dot P$               &      $2.299(38)\times 10^{-17}$     \\
 $\chi^2_{\nu}[{\rm DoF}]$                 &       2.32[13]                     \\
\cutinhead{Post-2010 Glitch Timing Solution (2011-2014) \hfill} 
 Epoch of ephemeris (MJD TDB)$^{\rm b}$      &   56359.00000177           \\
 Span of ephemeris (MJD)                   &      55890--56829                   \\
 Frequency, $f$                       &      2.35776347415(36) s$^{-1}$            \\
 Frequency derivative, $\dot f$  &    $-1.03(27)\times 10^{-16}$ s$^{-2}$   \\
 Period, $P$                           &      0.424130753981(64) s            \\
 Period derivative, $\dot P$               &      $1.85(49)\times 10^{-17}$      \\
 $\chi^2_{\nu}[{\rm DoF}]$                 &       0.26[9]                      \\
\noalign{\vskip 0.5em}\hline\noalign{\vskip 0.5em}
Glitch epoch (MJD)$^{\rm b}$                  &       55509(36)                     \\
 $\Delta f$\                        &      $2.13(36)\times 10^{ -9}$ s$^{-1}$    \\
 $\Delta f /f_{\rm pred}$                         &      $9(2)\times 10^{-10}$          \\
\cutinhead{Post-2015 Glitch Timing Solution (2016-2019) \hfill} 
 Epoch of ephemeris (MJD TDB)              &          58144.00000219           \\
 Span of ephemeris (MJD)                   &      57597--58695                   \\
 Frequency, $f$                     &      2.35776345915(16) s$^{-1}$ \\
 Frequency derivative, $\dot f$  &      $-1.01(12)\times 10^{-16}$ s$^{-2}$   \\
 Period, $P$                          &      0.424130756679(30) s           \\
 Period derivative, $\dot P$               &      $1.81(21)\times 10^{-17}$      \\
 $\chi^2_{\nu}[{\rm DoF}]$                 &       0.91[17]                     \\
\noalign{\vskip 0.5em}\hline\noalign{\vskip 0.5em}
 Glitch epoch (MJD)$^{\rm b}$                &      56800(8)                       \\
 $\Delta f$\                     &      $9(2)\times 10^{-10}$ s$^{-1}$         \\
 $\Delta f/f_{\rm pred}$                         &      $3.70(66)\times 10^{-10}$      \\[-0.5em]
\enddata
\tablenotetext{a}{Uncertainties in the last digits are given in parentheses.}
\tablenotetext{b}{Epoch of the glitch estimated by matching the zero phase 
of the two timing solutions; this assumes a constant post-glitch $\dot f$.}
\label{tab:ephem}
\end{deluxetable}

\begin{figure}
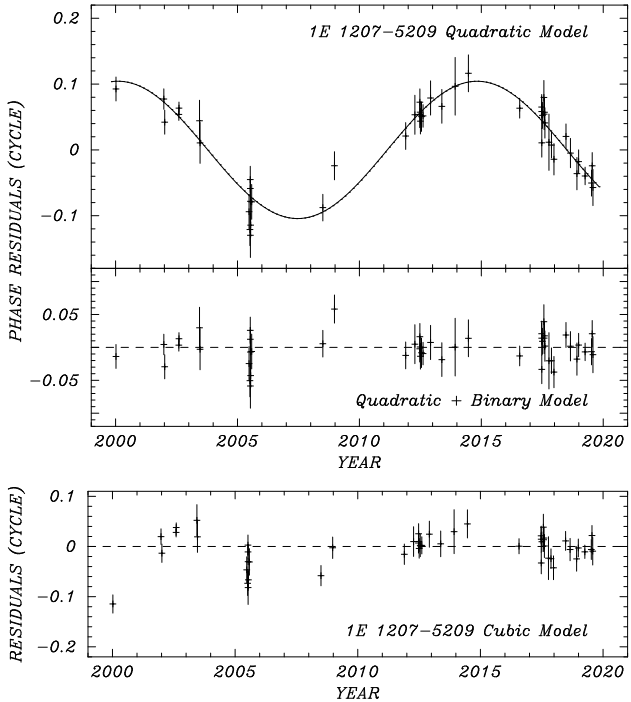

\centerline{\psfig{figure=1e1207_2000_2019_binary.ps,height=0.96\linewidth,angle=270}}
\vspace{2mm}
\centerline{\psfig{figure=1e1207_2000_2019_2_deriv.ps,height=0.96\linewidth,angle=270}}
\caption{ \footnotesize Pulse-phase residuals from the alternative
  timing models presented in Table~\ref{tab:ephemalt} that do not
  involve glitch(es).  These fits include the
  2000 \chandra\ data point (see Section~\ref{sec:three} for details).
  Top: A quadratic fit leaves sinusoidal oscillations in the phase
  residuals (upper panel).  The lower panel shows the residuals
  after a binary orbit having a period of $14.77\pm0.60$~yr and
  projected semimajor axis of $0.049\pm0.010$~lt-s is added.
  Bottom: Residuals from a cubic polynomial fit (including frequency
  second derivative).
}
\label{fig:altres}
\end{figure}

\begin{deluxetable}{lc}
\tighten
\tablewidth{0pt}
\tablecolumns{2}
\tablecaption{Alternative Timing Solutions for \cco}
\tablehead{
\colhead{Parameter} & \colhead{Value$^{\rm a}$ \hfill}
}
\startdata
\cutinhead{Quadratic + Binary Timing Solution (2000-2019)\hfill}
 Epoch of ephemeris (MJD TDB)           & 55478.00000457\\                                                                                
 Span of ephemeris (MJD)                    & 51549--58695\\
 Frequency, $f$                       & 2.357763483701(46) s$^{-1}$ \\
 Frequency derivative, $\dot f$  & $-1.1215(39) \times 10^{-16}$ s$^{-2}$ \\
 Period, $P$                           & 0.4241307522627(84) s \\
 Period derivative, $\dot P$                & $2.0176(71)\times 10^{-17}$ \\
 Binary period                          & $14.77(60)$ yr \\
 Projected semi-major axis              & $0.049(10)$ lt-s \\
 Time of ascending node (MJD)          & $55611(57)$ \\
 Longitude of periastron passage       & $342^{\circ}\pm4^{\circ}$ \\
 $\chi^2_{\nu}[{\rm DoF}]$                  & 1.13[43] \\
\cutinhead{Cubic Timing Solution (2000-2019) \hfill} 
 Epoch of ephemeris (MJD TDB)              &          55478.00000457           \\
 Span of ephemeris (MJD)                   &      51549--58695                   \\
 Frequency, $f$                      &      2.357763483047(42) s$^{-1}$           \\
 Frequency derivative, $\dot f$ &      $-1.1193(20)\times 10^{-16}$ s$^{-2}$  \\
 Frequency second derivative, $\ddot f$ & $6.57(39)\times 10^{-26}$ s$^{-3}$    \\
 Period, $P$                      &      0.4241307523805(76) s           \\
 Period derivative, $\dot P$               &      $2.0135(36)\times 10^{-17}$    \\
 Period second derivative, $\ddot P$  &   $-1.181(70)\times 10^{-26}$ s$^{-1}$   \\
 $\chi^2_{\nu}[{\rm DoF}]$                 &       3.04[45]                     \\
\cutinhead{Cubic Timing Solution (2002-2019) \hfill} 
 Epoch of ephemeris (MJD TDB)              &         55478.00000451           \\
 Span of ephemeris (MJD)                   &      52266--58695                   \\
 Frequency, $f$                   &      2.357763482808(50) s$^{-1}$            \\
 Frequency derivative, $\dot f$  &   $-1.1255(21)\times10^{-16}$ s$^{-2}$  \\
 Frequency second derivative, $\ddot f$    &      $9.18(49)\time 10^{-26}$ s$^{-3}$    \\
 Period, $P$                            &      0.4241307524235(90) s           \\
 Period derivative, $\dot P$               &      $2.0246(38)\times 10^{-17}$    \\
 Period second derivative, $\ddot P$      &      $-1.651(88)\times 10^{-26}$ s$^{-1}$   \\
 $\chi^2_{\nu}[{\rm DoF}]$                 &       1.34[44]                     \\[-0.5em]
\enddata
\tablenotetext{a}{Uncertainties in the last digits are given in parentheses.}
\label{tab:ephemalt}
\end{deluxetable}

\section {Timing Analysis}
\label{sec:timing}

For each reprocessed, cleaned event file, we folded the extracted
photon arrival times on the pulsar frequency to compute the
time-of-arrival (ToA) of phase zero of the pulse. To attempt a
phase-connected timing solution, we fit the set of ToAs using the
{\tt TEMPO} software \citep{hob06} to a model for the rotation phase
of the pulsar including one or two of its frequency derivatives,
$$ \phi(t) = \phi_{o} + f(t-t_{o}) + {1\over2}\dot f (t-t_{o})^{2}
+ {1\over6}\ddot f (t-t_{o})^{3}.$$
Initially, we obtained the fold
frequency from a periodogram search for the maximum power
around the expected frequency,
determined from the ephemeris of \cite{got18}. After
fitting these ToAs to generate  an intermediate timing solution, we iterated a
refined set of ToAs by folding the arrival times on the frequencies
predicted by this solution, before and after the glitch.

The summed pulse profile, generated by folding all the data together,
is found to be well-characterized by a sine function.  We use this
model to determine the phase zero for the ToAs, most accurately
computed from the unbinned photon arrival times, $t_{i}$, from the
ratio of the Fourier sums,
$$\psi_{\rm sine} = {\rm tan}^{-1} \Big\{
\sum\sin[2\pi\phi(t_i)]\Big/\sum\cos[2\pi\phi(t_i)] \Big\}.$$
The uncertainty in the phase is determined from a least-squares fit of a
sine function to the pulse profile folded in 20 phase bins.
In this work, phase zero is defined as the minimum of the modeled
sine.

\subsection {Single Glitch Fit}
\label{sec:one}

Figure~\ref{fig:oneres} graphs the ToA phase residuals from the
pre-glitch timing solution given in Table~\ref{tab:ephemone}, obtained
using data points from 2002--2015. 
After 2015, a linear deviation from this solution is evident and its slope
gives a change of frequency of $\Delta f = (5.03\pm0.16)\times
10^{-9}$~Hz and a glitch magnitude of $\Delta f / f_{\rm pred} =
(2.134\pm0.066)\times10^{-9}$.  The predicted frequency $f_{\rm pred}$ is
found by extrapolating the pre-glitch solution to the glitch epoch of
2014 November~21 (MJD~56982), estimated by matching the zero phase of the
pre- and post glitch solutions. The linearity of the post-glitch line
is consistent with a simple glitch in frequency; there is no evidence of
a change in the frequency derivative as suggested in \citet{got18}.
If there is any short-term partial recovery after the glitch,
it is not resolved by these sparse data.
The magnitude of the glitch is also about half of the value in
\citet{got18}.  Nevertheless, this is the same timing solution
as the one published previously, in the sense that
the cycle count calculated over the data span common to both
analyses is the same.  The parameters of the revised post-glitch
timing solution are simply made more accurate by including the
new data.  For the entire data set, $\chi^2_{\nu} = 1.44$ for 42 DoF,
taking into account the fit parameters for each interval.
We also note that the very first ToA, the
\chandra\ observation of 2000, does not seem to fit with
the pre-glitch analysis, so we ultimately excluded it from the
fits for the pre-glitch ephemeris in Table~\ref{tab:ephemone}.
This data point is nevertheless shown for reference
in Figure~\ref{fig:oneres} and in subsequent residual graphs.

\subsection {Two Glitch Fit}
\label{sec:two}

When extrapolated back to earlier times, the updated post-glitch
timing solution is sufficiently well sampled to reveal deviations in
the residuals that suggest an earlier glitch likely occurred
around epoch 2010. Figure~\ref{fig:postres} presents the residuals
from a 2002--2010 pre-glitch phase-connected solution extrapolated
to later times. A clear transition occurred at the estimated epoch
2010 November~9 (MJD~55509) with $\Delta f = (2.13\pm0.36)\times
10^{-9}$~Hz and $\Delta f / f_{\rm pred} = (9\pm2)\times10^{-10}$, about
half the value for the single glitch fit. Comparing the residual of
the second post-glitch solution to the first post-glitch solution,
we now measure a smaller glitch that occurred at 2014 May 23 (MJD 56800),
with $\Delta f = (9\pm2)\times 10^{-10}$~Hz and $\Delta f / f_{\rm pred} =
(3.70\pm0.66)\times10^{-10}$. Thus, there is now evidence of at least three
distinct spin-down intervals whose ephemerides are given in
Table~\ref{tab:ephem}.  The entire data set fitted with this model has
$\chi^2_{\nu} = 1.23$ for 39 degrees DoF, taking into account the fit
parameters for each interval.  

The timing models presented above are unable to
fit the year 2000 data point to within the uncertainty of the ToA.
This particular observation is hard to dismiss since it yields a highly
significant detection of the pulsed signal, and a high-quality ToA.
Furthermore, we have found no evidence for systematic error
associated with the data reduction or measurement of the ToA.  
If there was a glitch between 2000 and 2002, it could explain the discrepant
point in 2000.  This would suggest that glitches in \cco\
occur at an interval of $4-10$ years.

\subsection {Alternative Model Fits}
\label{sec:three}

Given the uncertain physics of CCO pulsars in particular
and glitches in general, we also tested alternative timing models
for the full data set that might fit without using glitches.
In such models there is no particular justification for excluding the
2000 \chandra\ data point, so we include it.
The timing solutions for these alternative models are given in
Table~\ref{tab:ephemalt}.

Starting with a simple
quadratic timing model (with one frequency derivative) leaves a sinusoidal
oscillation in the phase residuals.  As shown in Figure~\ref{fig:altres},
these residuals can be fully accounted for by a circular binary orbit
with a period of $14.77\pm0.60$~yr and a projected semi-major axis
of $0.049\pm0.010$~lt-s (Table~\ref{tab:ephemalt}).  The fit,
with $\chi^2_{\nu} = 1.13$ for 43 DoF, is as good as
or better than the glitch models.  We discuss the
possible interpretation of these fitted parameters in
Section~\ref{sec:noise}.

We also consider a cubic polynomial, which includes a frequency
second derivative, over the entire span of the observations.
This can also fully model the set of ToAs, but only if the 2000
point is excluded (Figure~\ref{fig:altres}, bottom 
panel).   Without the 2000 point, $\chi^2_{\nu} = 1.34$ for 44 DoF,
but including it yields $\chi^2_{\nu} = 3.04$ for 45 DoF.
Parameters from both versions of the cubic fit are given
in Table~\ref{tab:ephemalt}.  The frequency second derivative
$\ddot f$, or the braking index, defined as
$n \equiv f\ddot f/\dot f^2$, are
often used to characterize timing noise in pulsars.
In Section~\ref{sec:noise}
we discuss the cubic fit in terms of timing noise.

\section {Discussion}
\label{sec:discussion}

\subsection {Glitch Models}

Taking the glitch timing models at face value,
continued observations of \cco\ show that the previously
discovered glitch is better described as two smaller ones separated
by 3.5 yr.  In addition, there is no longer any evidence for a
large change in frequency derivative as suggested by \citet{got18}.
Nevertheless, the mere detection of glitching activity in a pulsar
with such a small spin-down rate is unprecedented, as we shall
describe below.  As shown most recently by
\citet{esp11} and \citet{fue17}, glitch activity is best
correlated with $\dot f$, such that $\approx1\%$ of the long-term
spin-down is reversed by glitching.  In the context of the
vortex creep theory of glitches \citep{alp84}, this implies
that $1\%$ or more of the moment of inertia of the
NS is contained in a crustal superfluid whose vortices are
repeatedly pinned and unpinned.

The glitch activity parameter for an
individual pulsar is defined as
$$\dot f_g \equiv {\sum_j \Delta f_j \over T},$$
where the numerator is the sum of the changes in frequency
over the glitches, and $T$ is the total span of the observations.
The linear correlation in which
$\dot f_g \approx 0.01 |\dot f|$ only becomes apparent when
glitch activity is summed over groups of pulsars binned in
$\dot f$.  However, this linear correlation holds only
in the range $10^{-14} < |\dot f| < 10^{-11}$ s$^{-2}$.
In addition, $f_g$ is dominated in this range by
the largest glitches, which have $\Delta f/f \sim 10^{-6}$
For smaller values of $|\dot f|$, only small glitches occur,
and glitch activity plummets such that no pulsar with
$|\dot f|<3\times10^{-16}$~s$^{-2}$ has been observed to glitch
in $1780$ pulsar years of monitoring \citep{fue17}.
The upper limit on the glitch parameter for such small $|\dot f|$
is $\dot f_g < 10^{-19}$ s$^{-2}$ by extrapolation
from pulsars with larger $|\dot f|$.

In contradistinction, \cco\ with its
$\dot f=-1.2\times10^{-16}$~s$^{-2}$ has glitched two or three times
in 20 years, with a glitch activity parameter of
$\dot f_g = (5-9)\times10^{-18}$ s$^{-2}$, which has the
result of reversing $\approx 4-7\%$ of its spin-down.
Evidently \cco\ experiences higher glitch activity relative
to its spin-down rate than most pulsars, its activity
being dominated by small but frequent glitches.

\subsection {Alternative Models}
\label{sec:noise}

Alternative models without glitches are equally good in fitting the
timing data on \cco.  In particular, the fit of quadratic spin-down
plus binary orbit has the lowest $\chi^2$ of the all of the models
tested here.  If the sinusoidal component is due to an orbital
motion, the minimum mass of the companion would be $6.8\,M_{\oplus}$
for a $1.4\,M_{\odot}$ NS, which is similar to the original pulsar
planets PSR B1257+12 B and C \citep{kon03}, albeit with a much longer
period of $\approx 15$~yr instead of $2-3$ months.  But the fitted
period, only slightly shorter than the time span of the observations,
is a typical result of red noise, a known characteristic of the timing
noise of pulsars, and thus a more likely interpretation.  In the
following, we quantify the timing noise and compare it with trends in
the general pulsar population.

Several diagnostics of timing noise have been introduced over
the years.  They were reviewed recently by \citet{nam19} in
their study of timing noise in 129 middle-aged pulsars,
and we employ three of the methods here.  First is a simple
metric favored by \citet{sha10},
$$\sigma_{\rm TN,2}^2 = \sigma_{\rm R,2}^2 - \sigma_{\rm W}^2,$$
where $\sigma_{\rm R,2}$ is the root-mean-square (rms) of the measured residuals
from a second-order polynomial fit, and $\sigma_{\rm W}$
is the typical uncertainty of a ToA.  (The subscripts R and W
refer to red and white noise processes, respectively.)
\citet{sha10} found for hundreds of canonical pulsars
(not millisecond pulsars or magnetars)
that the mean value of $\sigma_{\rm TN,2}$ scales with the
spin parameters as
$$\bar\sigma_{\rm TN,2} = C_2\,f^{\alpha}\,|\dot f|^{\beta}\,T^{\gamma}
\ \ \mu{\rm s},$$
where $C_2 = 41.7, \alpha=-0.9, \beta=1.00$, and $\gamma=1.9$.
Recognizing that there is large scatter in $\sigma_{\rm TN,2}$,
\citet{sha10} modelled the distribution as log-normal,
and found that the standard deviation of log($\sigma_{\rm TN,2}$)
is $\delta = 1.6$ 

This method is applicable to the quadratic fit
of Figure~\ref{fig:altres} (top).
The rms timing residual is $\sigma_{\rm R,2}=27.3$~ms, while
the average uncertainty of a ToA is $\sigma_{\rm W}=9.5$~ms;
therefore, $\sigma_{\rm TN,2}=25.6$~ms.  In comparison,
the fitted value of $\bar\sigma_{\rm TN,2}$ from \citet{sha10}
corresponding to the timing parameters of \cco\ is
$\approx100\,\mu$s (see also the data in Figure~6
of \citealt{nam19}).  The observed residuals therefore
exceed the pulsar average by a factor of $\approx250$,
which is much greater than the scatter of
$10^{1.6}$ found by \citet{sha10} and the
scatter of the data points in \citet{nam19}.  This
shows that, if the timing irregularities in \cco\
are timing noise, it is behaving like a pulsar with
2--3 orders of magnitude larger $|\dot f|$ or $B_s$.

An earlier parameterization of timing noise is that of
\citet{arz94}, who used the frequency second derivative
measured over a time span of $T=10^8$~s to define
$$\Delta_8 = {\rm log} \left({1\over6f}\,|\ddot f|\,T^3\right).$$
\citet{arz94}, \citet{hob10}, and
\citet{nam19} showed that $\Delta_8$ is
positively correlated with $\dot f$.
Unfortunately, this and other metrics are sensitive to the time
span of the observation, since $\ddot f$ itself,
being the result of red noise, generally
increases with $T$.  Therefore,
$\Delta_8$ should not be used to compare pulsars over
different time spans.  But since the ToAs of \cco\ are not
nearly as precise as those of radio pulsars, we cannot even
get a significant measurement of $\ddot f$ if we reduce the
time span of the fit to $10^8$~s.  Acknowledging the
limitations of such a comparison, we nevertheless
calculate $\Delta_8 = 0.04$ from either version of
the cubic fit in Table~\ref{tab:ephemalt}, finding that it
is 2--3 orders of magnitude larger than that
of pulsars with similar $\dot f$, and at the high end
of all pulsars.  This rather extreme discrepancy argues
that \cco\ is much noiser than pulsars with similar
spin-down rates.

Finally, the braking index itself can be used to characterize
timing noise.  For the timing parameters of the two cubic fits
in Table~\ref{tab:ephemalt}, $n = 1.2\times10^7$ or
$1.7\times10^7$, which are off the scale of values
plotted in \citet{nam19}.  Pulsars of similar $\dot f$ or $B_s$
have $n<2\times10^6$.

\section {Conclusions}
\label{sec:conclusions}

Whether the timing irregularities of \cco\ are described as
two or three glitches, or as timing noise,
the magnitude of the effects are much greater than
in radio pulsars of similar spin-down rate or
dipole magnetic field strength.  Because \cco\ displays
no evidence of magnetospheric activity, while its timing properties
are commensurate with those of the young pulsar population,
an internal property such as high temperature or high
$B$-field strength is implicated.
\cite{ho15} proposed that glitches could be triggered
by the motion of magnetic fields through the NS crust,
interacting with the neutron superfluid there.  If so,
a magnetic field much stronger than the surface dipole
field is buried in the crust of \cco. \citet{ho15}
was envisioning large glitches like those of the Vela
pulsar, whereas only small glitches have so far been
detected from \cco, which distinguishes it from the more
energetic pulsars.

Timing noise has been attributed to variability in the
interaction of the crustal superfluid with the Coulomb
lattice of the solid crust \citep{jon90}, turbulence
of the superfluid \citep{mel14}, or fluctuations in
the structure of the magnetosphere, e.g., 
state switching \citep{lyn10}.
Just as for glitches, internal effects would be favored
as the cause of timing noise in \cco\ because of its lack
of magnetospheric activity.

Finally, as discussed in \citet{got18}, it has not been ruled
out that low-level accretion from an undetected fall-back
debris disk could be the cause of its timing fluctuations while
making a negligible contribution to the luminosity of \cco.
The present results do not alter those arguments about
accretion torques, and we do not repeat them here, except
to recall the possible connection between field burial and
formation of a  residual disk, which would require
only a small fraction of the fall-back debris to be held
in reserve for long-term accretion from a disk.

\acknowledgments
Support for this work was provided by NASA through {\it XMM\/} grants 80NSSC18K0452, 80NSSC19K0866,
\nicer\ grant 80NSSC19K1461, and {\it Chandra} Award SAO GO7-18063X issued by the {\it Chandra} X-ray Observatory Center, which is operated by the Smithsonian Astrophysical Observatory for and on behalf of NASA under contract NAS8-03060. This investigation is based partly on observations obtained with \xmm, an ESA science mission with instruments and contributions directly funded by ESA Member States and NASA. This research has made use of data and/or software provided by the High Energy Astrophysics Science Archive Research Center (HEASARC), which is a service of the Astrophysics Science Division at NASA/GSFC.


\begin{thebibliography}{}

\bibitem[Alpar \etal(1984)]{alp84}
Alpar, M. A., Pines, D., Anderson, P. W., \& Shaham, J. 1984,
\apj, 276, 325

\bibitem[Arzoumanian \etal(1994)]{arz94}
Arzoumanian, Z., Nice, D. J., Taylor, J. H., \& Thorsett, S. E. 1994,
\apj, 422, 671

\bibitem[Bignami \etal(2003)]{big03}
Bignami, G. F., Caraveo, P. A., De Luca, A., \& Mereghetti, S. 2003, Natur, 423, 725

\bibitem[De~Luca \etal(2004)]{del04}
De Luca, A., Mereghetti, S., Caraveo, P. A., \etal\ 2004, A\&A, 418, 625 

\bibitem[De~Luca \etal(2017)]{del17}
De Luca, A. 2017, JPhCS, 932, 012006 

\bibitem[Espinoza \etal(2011)]{esp11}
Espinoza, C. M., Lyne, A. G. Stappers, B. W., \& Kramer, M. 2011,
\mnras, 414, 1679

\bibitem[Fuentes \etal(2017)]{fue17}
Fuentes, J. R., Espinoza, C. M., Reisenegger, A., \etal\ 2017, \aap, 608, A131

\bibitem[Gendreau \& Arzoumanian(2017)]{gen17}
Gendreau, K., \& Arzoumanian, Z. 2017, NatAs, 1, 895

\bibitem[Gotthelf \& Halpern(2007)]{got07}
Gotthelf, E. V., \&  Halpern, J. P. 2007, ApJL, 664, L35

\bibitem[Gotthelf \& Halpern(2018)]{got18}
Gotthelf, E. V., \& Halpern, J. P. 2018, \apj, 866, 154

\bibitem[Gotthelf \etal(2013)]{got13}
Gotthelf, E. V.,  Halpern, J. P.,  Alford, J. 2013, ApJ, 765, 58
  
\bibitem[Halpern \& Gotthelf(2010)]{hal10}
Halpern, J. P., \& Gotthelf, E. V. 2010, ApJ, 709, 436

\bibitem[Halpern \& Gotthelf(2011)]{hal11}
Halpern, J. P., \& Gotthelf, E. V. 2011, ApJL, 733, L28

\bibitem[Halpern \& Gotthelf(2015)]{hal15}
Halpern, J. P., \& Gotthelf, E. V. 2015, \apj, 812, 61

\bibitem[Ho(2015)]{ho15} 
Ho, W. C. G. 2015, MNRAS, 452, 845

\bibitem[Hobbs \etal(2006)]{hob06}  
Hobbs, G. B., Edwards, R. T., \& Manchester, R. N. 2006, \mnras, 369, 655

\bibitem[Hobbs \etal(2010)]{hob10}  
Hobbs, G., Lyne, A. G., \& Kramer, M. 2010, \mnras, 402, 1027

\bibitem[Jones(1990)]{jon90}
Jones, P. B. 1990, \mnras, 246, 364
  
\bibitem[Konacki \& Wolszczan(2003)]{kon03}
Konacki, M. \& Wolszczan, A. 2003, ApJL, 591, L47
  
\bibitem[Lyne \etal(2010)]{lyn10}
  Lyne, A., Hobbs, G., Kramer, M., Stairs, I., \& Stappers, B. 2010,
  Sci, 329, 408

\bibitem[Melatos \& Link(2014)]{mel14}
Melatos, A., \& Link, B. 2014, \mnras, 437, 21

\bibitem[Mereghetti \etal(2002)]{mer02}
Mereghetti, S., De Luca, A., Caraveo, P. A., \etal\ 2002, \apj, 581, 1280

\bibitem[Namkham \etal(2019)]{nam19} Namkham, N., Jaroenjittichai, P.,
\& Johnston, S. 2019, \mnras, 487, 5854

\bibitem[Okajima \etal(2016)]{oka16}
Okajima, T., Soong, Y., Balsamo, E. R., \etal\ 2016, Proc. SPIE, 9905, 99054X


\bibitem[Pavlov \etal(2002)]{pav02}
Pavlov, G. G., Sanwal, D., Garmire, G. P., \& Zavlin, V. E. 2002,
in ASP Conf. Ser. 271, Neutron Stars in Supernova Remnants,
ed. P. O. Slane \& B. M. Gaensler (San Francisco, CA: ASP), 247

\bibitem[Prigozhin \etal(2016)]{pri16}
Prigozhin, G., Gendreau, K., Doty, J. P., \etal\ 2016, Proc. SPIE, 9905, 99051I

\bibitem[Sanwal \etal(2002)]{san02}
Sanwal, D., Pavlov, G. G., Zavlin, V. E., \& Teter, M. A. 2002, ApJL, 574, L61

\bibitem[Shannon \& Cordes(2010)]{sha10}
Shannon, R. M., \& Cordes, J. M. 2010, \apj, 725, 1607

\bibitem[Zavlin \etal(2000)]{zav00}
Zavlin, V. E., Pavlov, G. G., Sanwal, D., \& Tr\"umper, J. 2000,
ApJL, 540, L25

\end{thebibliography}
\end{document}